\documentclass[aps,prb,twocolumn,groupedaddress,showpacs]{revtex4-1}
		
		\usepackage{graphicx}
		\usepackage{amsmath}
        \usepackage[usenames,dvipsnames]{color}
        \usepackage{pdfcolmk}

\begin{document}

        \title{Splitting Valleys in Si/SiO$_2$: Identification and Control of  Interface States}

\author{Amintor Dusko}

\affiliation{Instituto de F\'{\i}sica, Universidade Federal do Rio de Janeiro, Caixa Postal 68528, 21941-972 Rio de Janeiro, Brazil}

\author{A. L. Saraiva}

\affiliation{Instituto de F\'{\i}sica, Universidade Federal do Rio de Janeiro, Caixa Postal 68528, 21941-972 Rio de Janeiro, Brazil}

\author{Belita Koiller}

\affiliation{Instituto de F\'{\i}sica, Universidade Federal do Rio de Janeiro, Caixa Postal 68528, 21941-972 Rio de Janeiro, Brazil}

		\date{\today}
		
		\begin{abstract}

Interface states in a silicon/barrier junction break the silicon valley degeneracy near the interface, a desirable feature for some Si quantum electronics applications. Within a minimal multivalley tight-binding model in one dimension, we inspect here the spatial extent of these states into the Si and the barrier materials, as well as favorable conditions for its spontaneous formation. Our approach---based on Green's-function  renormalization-decimation techniques---is asymptotically exact for the infinite chain and shows the formation of these states regardless of whether or not a confining electric field is applied.
The renormalization language naturally leads to the central role played by the chemical bond of the atoms immediately across the interface.
 In the adopted decimation procedure, the convergence rate to a fixed point
 directly relates the valley splitting and the spread of the wave function, consequently connecting the splitting to geometrical experimental parameters such as the capacitance of a two-dimensional electron gas---explicitly calculated here. This should serve as a probe to identify such states as a mechanism for enhanced valley splitting.

		\end{abstract}


		\maketitle
		
		\section{\label{sec:Introduction}Introduction}

While most aspects of the Si/insulator interface are well understood~\cite{ando1982}, the emergence of interface states still presents puzzles for the semiconductor community. For classical devices---whose operation involves a macroscopic number of electrons---most of the drawbacks introduced by interface states may be overcome by modern growth techniques lowering their density in comparison to conduction states.
These techniques do not tackle the issue of devices operating at the quantum regime of one or a few electrons, though.
Understanding these states, we might be able to circumvent eventual associated problems and possibly make them instrumental. For instance, it was demonstrated that these states may be used for nuclear spin readouts~\cite{dreher_2012}. Moreover, it is conjectured that these states strongly break the Si valley degeneracy, an important desideratum in many silicon-based quantum computer architectures.~\cite{vrijen_2000,culcer_2012}

The Si conduction band edge is sixfold degenerate, with valley minima at wave vectors
{\small \begin{equation}\label{CB_Minima}
\left\{{\mathbf k_{\mu}}\right\}_{\mu=\pm x,\pm y,\pm z} =\left\{(\pm k_0, 0, 0);(0,\pm k_0,0);(0,0,\pm k_0)\right\},
\end{equation}}
where $k_0=0.85 \left(2\pi/a_{\rm Si}\right)$, $a_{\rm Si}$ is the Si conventional lattice parameter. Aiming at the development of Si-based quantum devices, a deeper understanding and clearer identification of the mechanisms lifting this degeneracy constitute a key issue. While the potential of donor impurities, singular in 3D, successfully splits these states, leading to a non-degenerate orbital ground state, the picture is far more challenging at an interface, where the barrier potential is two dimensional. The mechanism of intervalley scattering by the barrier potential, which we assume to be in the [001] direction, efficiently splits the two $z$-valleys from the other four valleys, but the coupling between them induced by the abrupt interface giving the ground-state splitting (or valley splitting) is not as large as is desirable.
So far, theoretical estimates~\cite{Sham1979,ando1982,Boykin2004,saraiva2010,Culcer2010,Saraiva2011} and most experiments~\cite{ando1982,goswami2007} report relatively small (less than 1 meV) splitting of the lowest valley state.

In contrast, several works \cite{ouisse1998,takashina2004,takashina2006,niida2013} on Si/SiO$_2$ indicate that buried oxide silicon-on-insulator (BOX-SOI) interfaces may efficiently couple the $z$-valley states, leading to ground-state splitting orders of magnitude higher than those produced by regular thermal interfaces. One possible explanation for this effect \cite{saraiva2010} is related to the presence of interface states, which form spontaneously at some semiconductor-barrier interfaces and---in the context of Si-based classical and quantum electronics---may improve or hinder a given device's performance according to its functionality and the interface state properties.

In a recent study,~\cite{saraiva2010} intrinsic Si/SiO$_2$ interface states
and its hybridization to the Si bulk states were investigated, and it was
shown that this hybridization follows a valley selection rule, which
significantly enhances the ground-state splitting.
This mechanism was numerically investigated in Ref.~[\onlinecite{saraiva2010}]
within minimal {one and two dimensional} multivalley tight-binding models, both leading essentially to the same conclusions. It was thus conjectured that
this is the  prevailing mechanism leading to the giant valley splitting
observed in  BOX-SOI heterostructure samples.

The one-dimensional tight-binding model for the Si/barrier electronic structure is explored here from a different and more insightful perspective. An analytic (Green's function) formalism is developed from which the full range of parameter space is readily accessible and investigated. The microscopic physics of interface states emerges by approaching this problem within  a decimation technique based on renormalization-group ideas,~\cite{SILVA1981} which we generalize to account for second nearest neighbors, as required by our minimal multivalley tight-binding model.
Localized states are intrinsically related to the junction of two semi-infinite chains (modeling the Si and barrier material, respectively),
and undesirable effects\cite{saraiva2010} due to a finite-size supercell, periodic boundary conditions, or applied electric field are eliminated. Moreover, the renormalization procedure provides quantitative estimates for the localization lengths of the intrinsic interface states.\cite{Robbins1985} This information emerges from the decimation rate of convergence and is useful to experimentally probe and compare the participation of the interface states in the composition of the electronic states in different quantum devices. We suggest capacitance measurements around the junction region as a possible gauge to detect the presence of intrinsic interface states, differentiating those from the more usual interface states\cite{fang1966} bound to a triangular-shaped well -- formed by an electric field near a barrier interface.

\section{\label{sec:Formalism}Formalism}

The six-fold degeneracy in Si, ${\mu=\pm x,\pm y,\pm z}$ [see Eq. \ref{CB_Minima}], is partially lifted in the presence of an interface, which breaks the symmetry of the system along the direction perpendicular to it, $z$ here.
Assuming a perfectly flat interface, translational symmetry parallel to the $xy$ plane is preserved, while the potential profile along $z$ raises the energy of the $\mu={\pm x,\pm y}$ valleys with respect to the  $\mu={\pm z}$, and lifts the degeneracy of the resulting two-dimensional $\{\mathbf k_{\pm z}\}$  subspace. We therefore restrict our study to the $z$-direction, where the interface perturbation potential and interface localized state envelope are evident. As argued below, we restrict the barrier material to SiO$_2$.

\begin{figure}[h]
		\includegraphics[width=\columnwidth]{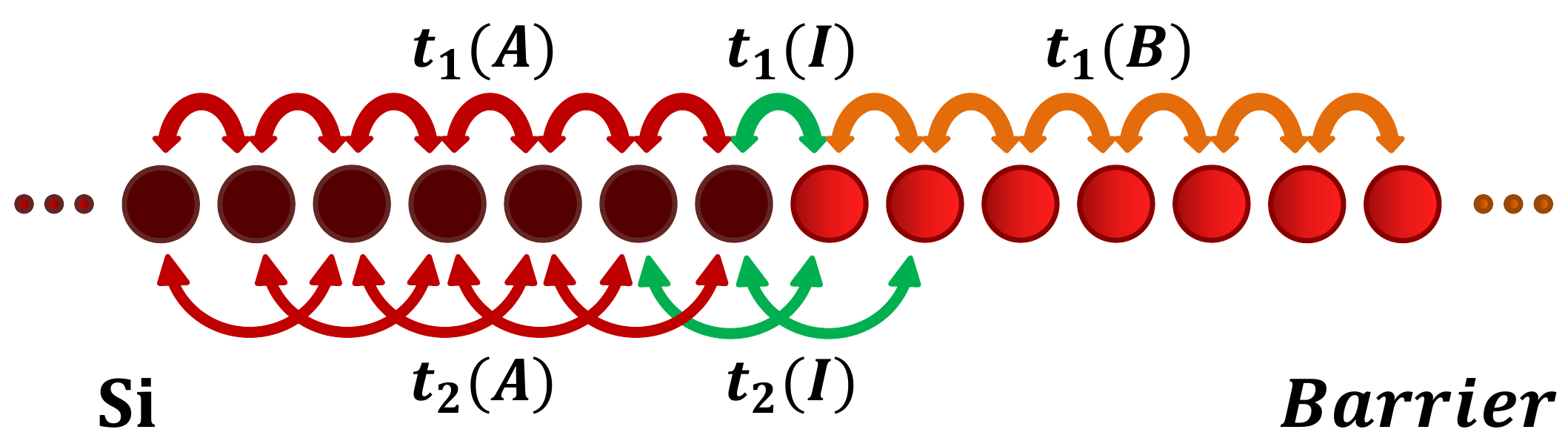}
		\caption{\label{fig:chain}(Color online) 
Schematic representation of the tight-binding one-dimensional model for the Si/barrier interface.
The labels $A$, $B$, and $I$ refer to Si, barrier, and interface regions, respectively. Dark sites are labeled by $s=-\infty,\ldots,-2, -1$, and represent Si occupation, while light sites indicate the effective barrier species, at  $s=1,2,\ldots, \infty$. The junction $A$ {\bf---} $B$ corresponds to bond $(-1)$ {\bf---} $(1)$, and the $I$ region includes the range $-2\le s \le 2$. The label $s=0$ is discarded so that $ A \longleftrightarrow B $ symmetry is obtained by changing $ -s \longleftrightarrow s $ along the decimation procedure (see Appendix \ref{app:decimation}).}
		\end{figure}

The one-dimensional model for the Si/SiO$_2$ (001) heterostructure consists of two connected semi-infinite chains  (light and dark sites in Fig.~{\ref{fig:chain}}) extending towards $\pm \infty$ away from the junction.
For the Si half-chain, we adopt a minimal one-orbital-per-site tight-binding
description of the conduction band accounting for the $z$-valleys physics.\cite{Boykin2004}

Modelling the SiO$_2$ layer realistically is not a trivial task within the empirical tight-binding approach, since this material is in general amorphous and its chemical bonds involve strong charge transfer. Nevertheless, the details of the electronic structure of the oxide have only a modest influence on the interface states \cite{saraiva2010}, so that we may choose a plausible arbitrary model without compromising the generality of our results.
For definiteness, we choose tight-binding parameters to fit the effective mass  of the $\beta$-cristobalite polymorph, and we adjust the on-site parameters so as to give a 3~eV conduction-band offset (barrier height).
For the description of this direct-gap structure, a tight-binding
parametrization with one-orbital per site and a nearest-neighbors hopping range
suffices, while first- and second-neighbors hopping links are needed to
account for the indirect gap of Si.

The effect of dimensionality, decisive in many physical contexts, is not as important for the properties investigated here.\cite{saraiva2010}
Even though the Van Hove singularities and the transport properties can only be correctly modelled within a full three-dimensional model, the valley splitting is determined only by the band profile across the (001) interface. This can be understood in terms of first-order perturbation theory, where the valley orbit integral between orthogonal valleys vanishes for a perfectly flat interface (see Ref.~[\onlinecite{Saraiva2011}]).

The complete set of parameters is given in Table {\ref{tab:parameters}}, and the nonzero hopping links are schematically represented in Fig.~{\ref{fig:chain}}. We take the unit cell to consist of two atomic sites, representing the two inequivalent consecutive planes of the diamond structure of Si---this choice has no particular physical meaning or effect for the barrier oxide. The (3D) lattice parameter for Si
(SiO$_2$) is $a_{\textrm{Si}}=0.54$~nm ($a_{\textrm{SiO}_2}=0.74$~nm). 
This leads to a distance between sites in the linear chain of $0.27$~nm ($0.37$~nm).
				
\begin{figure}[h!]
		\includegraphics[width=0.7\columnwidth]{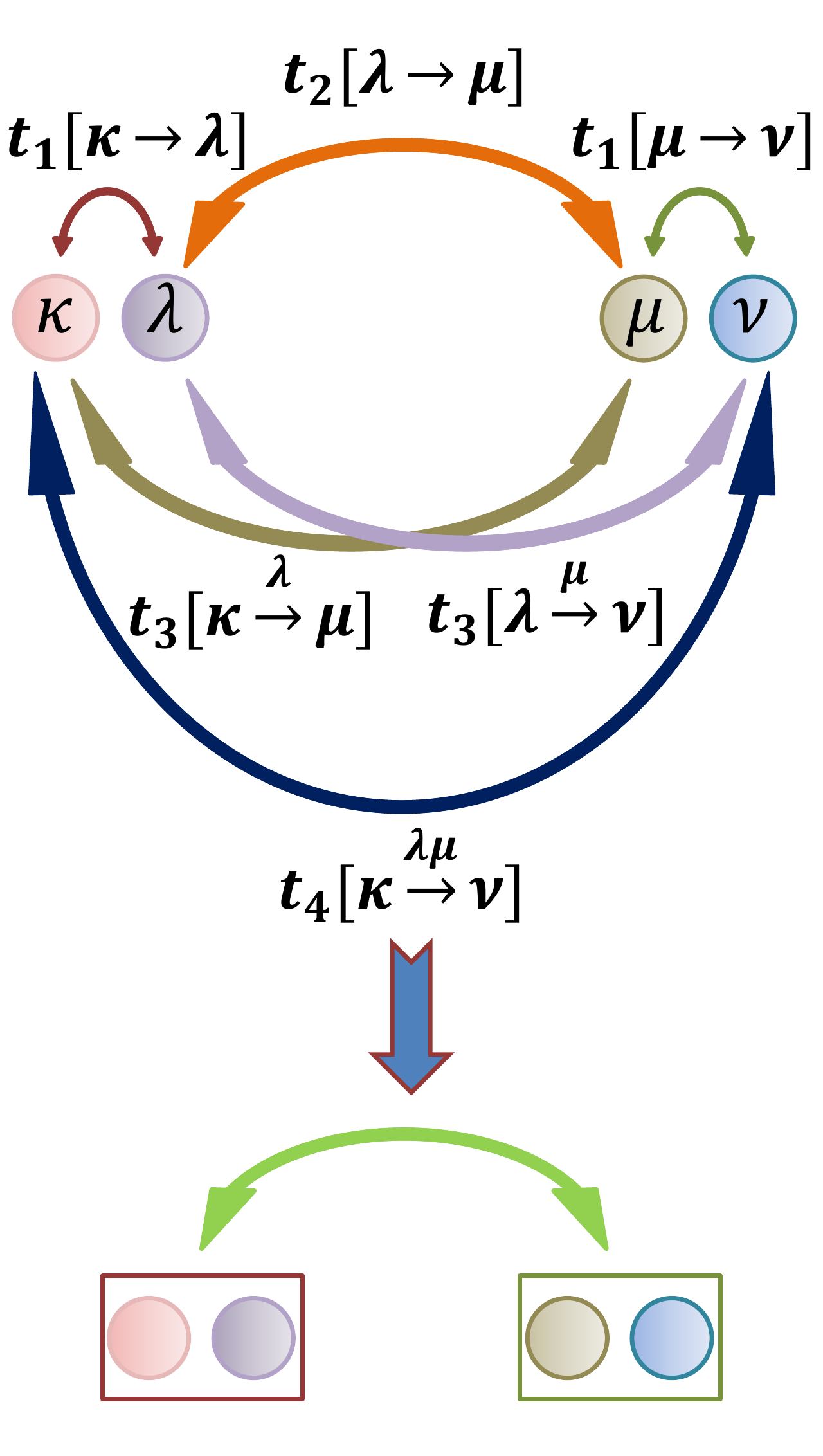}
		\caption{\label{fig:Dimers}(Color online) 
Diagrammatic structure of the renormalized hoppings explicitly appearing in the formalism after $n\geq 1$ decimation steps.
At each stage alternate dimers are projected out from Dyson's equations (see Appendix \ref{app:decimation}).
For the original chain with hoppings up to second nearest neighbors the decimated configurations involve hoppings ranging from first up to fourth neighbors.}               		
\end{figure}

\begin{figure}[h!]
		\includegraphics[width=\columnwidth]{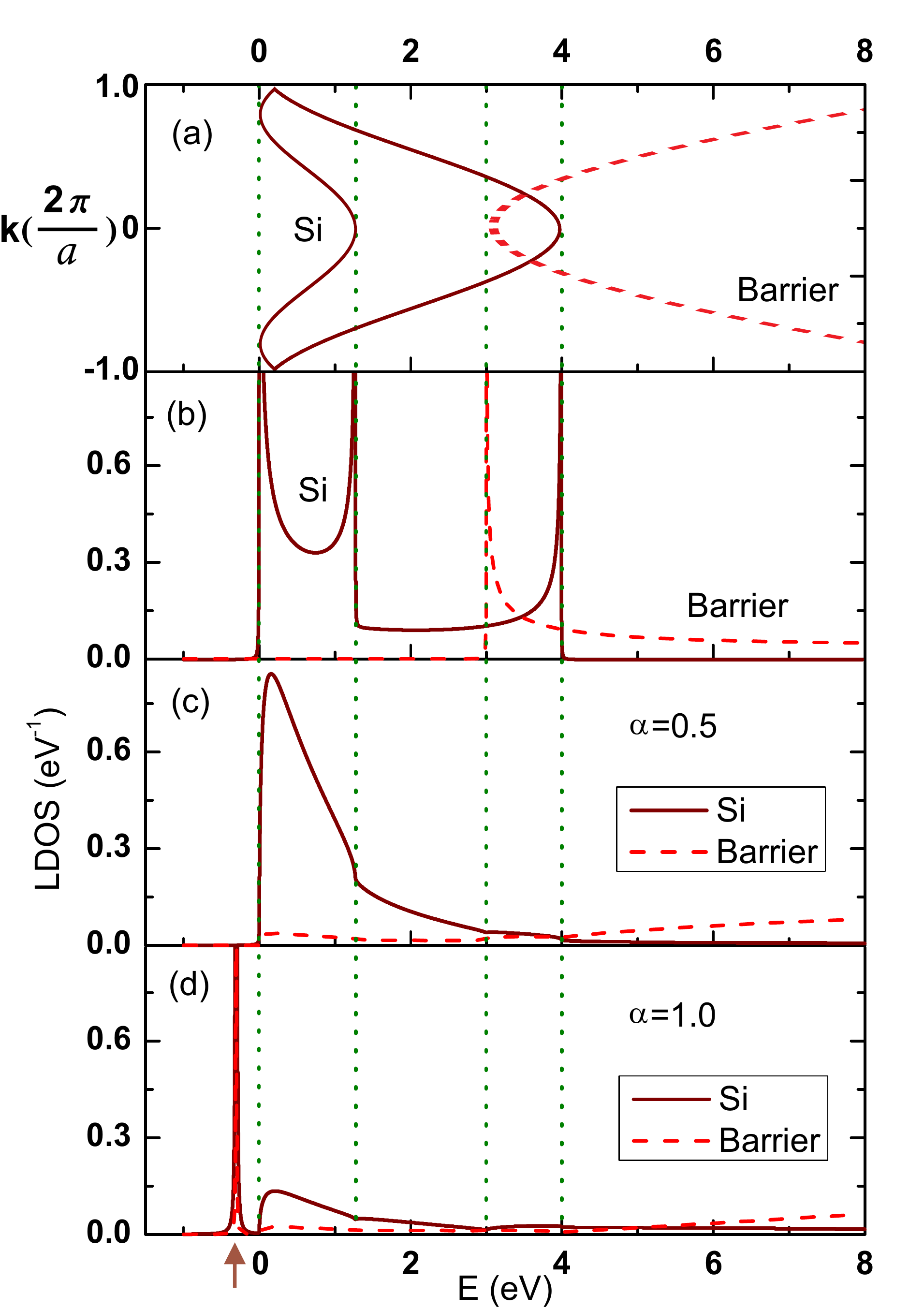}
		\caption{\label{fig:LDOS}(Color online) 
(a) Energy dispersion $E(k)$ for the bulk materials as indicated, with $a= a_{Si}$ ($a_{SiO_2}$) for the dispersion of Si (SiO$_2$).
(b) Local density of states (LDOS) in bulk Si and oxide,
(c) LDOS at the interface sites with $\alpha$=0.5 and
(d) LDOS at the interface sites with $\alpha$=1.0: here an interface state splits from the lower band into the gap, as indicated by the arrow.
The vertical dotted lines mark the Van Hove singularities of the one-dimensional tight-binding model for bulk Si and bulk oxide parameters.}
\end{figure}

The tight-binding Hamiltonian for the conduction band written in the basis set of one Wannier orbital $|s\rangle$ at each site $s$ is then

	\begin{widetext}	
		\begin{equation}
		\widehat{H}=\sideset{}{'}\sum_{s=-\infty}^{+\infty}\left\{ \epsilon_{s}|s \rangle\langle s|+\sum_{s_{nn}} t_{1}\left[\,s \xrightarrow{} {s_{nn}}\, \right]|s\rangle \langle s_{nn}|+\sum_{s_{nnn}}t_{2}\left[s \xrightarrow{j} {s_{nnn}}\right]|s\rangle \langle s_{nnn}|\right\}.
		\label{eq:Hamiltonian}
		\end{equation}
	\end{widetext}
Here $\epsilon_{s}$ is the diagonal term at site $s$, $t_{1}$ and $t_{2}$ are first and second neighbors tunnel couplings (or hopping parameters) and the first summation is primed to exclude $s=0$ {(See Fig.~{\ref{fig:chain}})}. The two first- (second-) nearest neighbors of $s$ are $s_{nn} (s_{nnn})$. Without ambiguity, the notation $s$ gives not only the sites sequence, but also the species and environment of site $s$: $s<-2 \to A (Si),~s>2 \to B$ (barrier)~or~$-2\leq s\leq2 \to I $ (interface) according to its location with respect to the junction   $(-1)A$ {\bf ---} $(+1)B$. The common first neighbor between $s$ and $s_{nnn}$ is labeled $j$, and  the notation  {$t_2\left[s \xrightarrow{j} s_{nnn}\right]$} indicates that this second-nearest-neighbors hopping is across a $j$-species site. (See Fig. \ref{fig:Dimers} ).

Since the $I$ couplings are not {\it a priori} determined, a variable $\alpha$ is introduced which linearly interpolates~\cite{saraiva2010} the first- and second-neighbors off-diagonal terms $t_1$ and $t_2$ from the Si tight-binding parameters $\left(\alpha=0\right)$ to the oxide parameters $\left(\alpha=1\right)$, as detailed in Table \ref{tab:parameters}.

		\begin{table} [h]
 \caption{\label{tab:parameters}
Tight-binding parameters, in eV, for the adopted one-dimensional model. The on-site energy is $\epsilon$ and $t_1$ $(t_2)$ is the nearest- (next-nearest-) neighbor hopping parameter. The labels $A$, $B$ and $I$ refer to Si, barrier, and interface regions, respectively, and $\alpha\in\left[0,1\right ]$.}
	\begin{ruledtabular}
		\begin{tabular}{l l c }
	   & $\epsilon_{A}$  &  1.41 \\
	Region A - Si & $t_{1}(A)$ &-0.68 \\
	  & $t_{2}(A)$ & 0.61 \\ \\
	Region I - Interface  & $t_{1}(I)$&$(1-\alpha)t_{1}(A)+\alpha t_{1}(B)$   \\
      & $t_{2}(I)$&$(1-\alpha)t_{2}(A)+\alpha t_{2}(B)$   \\

	\\
	     & $\epsilon_{B}$ & 9.56  \\
	Region B - Barrier & $t_{1}(B)$ & -3.28 \\
	  & $t_{2}(B)$ & 0 \\
		\end{tabular}
	\end{ruledtabular}
\end{table}
The density of states is obtained here from the Green's function $\widehat{G}(Z) = (Z .\widehat{\textbf{1}}  -\widehat{H})^{-1}$, with  $Z=E+i~u$, where $u$ is a small imaginary part added to the energy $E$ to avoid singularities of the Green operator and $\widehat{\textbf{1}}$ the unitary operator. Results presented here are for $u=10^{-3}$ meV.  The matrix elements $G_{ij}=\langle i|\widehat G |j\rangle $ for the  Hamiltonian in (\ref{eq:Hamiltonian}) are obtained exactly by a decimation approach,\cite{SILVA1981,Robbins1983} adapted here as  detailed in the Appendix \ref{app:decimation}.

The energetics of these states is given intrinsically by the bonds across the
interface. As shown in the Appendix \ref{app:decimation}, the decimation technique conveys this message in a clear and simple picture, by mapping the electronic Hamiltonian of the two semi-infinite chains into a fictitious molecule of two effective atomic species with renormalized self-energies (each representing one of the materials) connected by a single renormalized hopping.

The local electronic density of states at site $j$ and energy $E$, $LDOS(s=j, E)=-\left(1/\pi\right) \lim_{u\to 0}\Im \left[G_{jj}(Z)\right]$ and the total
density of states  $DOS(E)=\lim_{N\to \infty}\sum_{j=-N/2,N/2}[LDOS(s=j,E)/N]$ are then expressed in terms of the diagonal matrix elements.
The LDOS of sites very far from the interface ($|s|\gg 1$) asymptotically
coincides with the LDOS of the bulk material, while interface effects appear
for $|s|\sim 1$. Results for $E(k)$ and the LDOS of the bulk materials are presented in Fig.~\ref{fig:LDOS}(a) and \ref{fig:LDOS}(b) respectively. Two of the singularities in Fig.~\ref{fig:LDOS}(b) for Si occur at the minimum and maximum of the conduction band (band edges); a third peak is due to a local maximum at the direct gap ($k=0$) energy [see Fig.~\ref{fig:LDOS}(a)]. For the barrier, the singularity in Fig.~\ref{fig:LDOS}(b)corresponds to the lower conduction band edge ($k=0$) [see Fig.~\ref{fig:LDOS}(a)]. The singularity at the band maximum is outside of the energy range here.

\section{\label{sec:InterfaceState}Interface state energy and localization properties}

Intrinsic interface states were originally considered by Tamm \cite{Tamm1932} and Shockley \cite{Shockley1939}, and we refer to such states here as TS states.
The signature of a TS state at energy $E_{\textrm{TS}}$ is a pole in $G_{jj}(E)$ at
$E_{\textrm{TS}}$, which contributes to LDOS $(s=j)$ at sites $j$ close enough to the
interface with a $\delta$-function peak. The imaginary part $u$
added to $E$ broadens such peaks into Lorentzians.

\begin{figure}[h!]
		\includegraphics[width=\columnwidth]{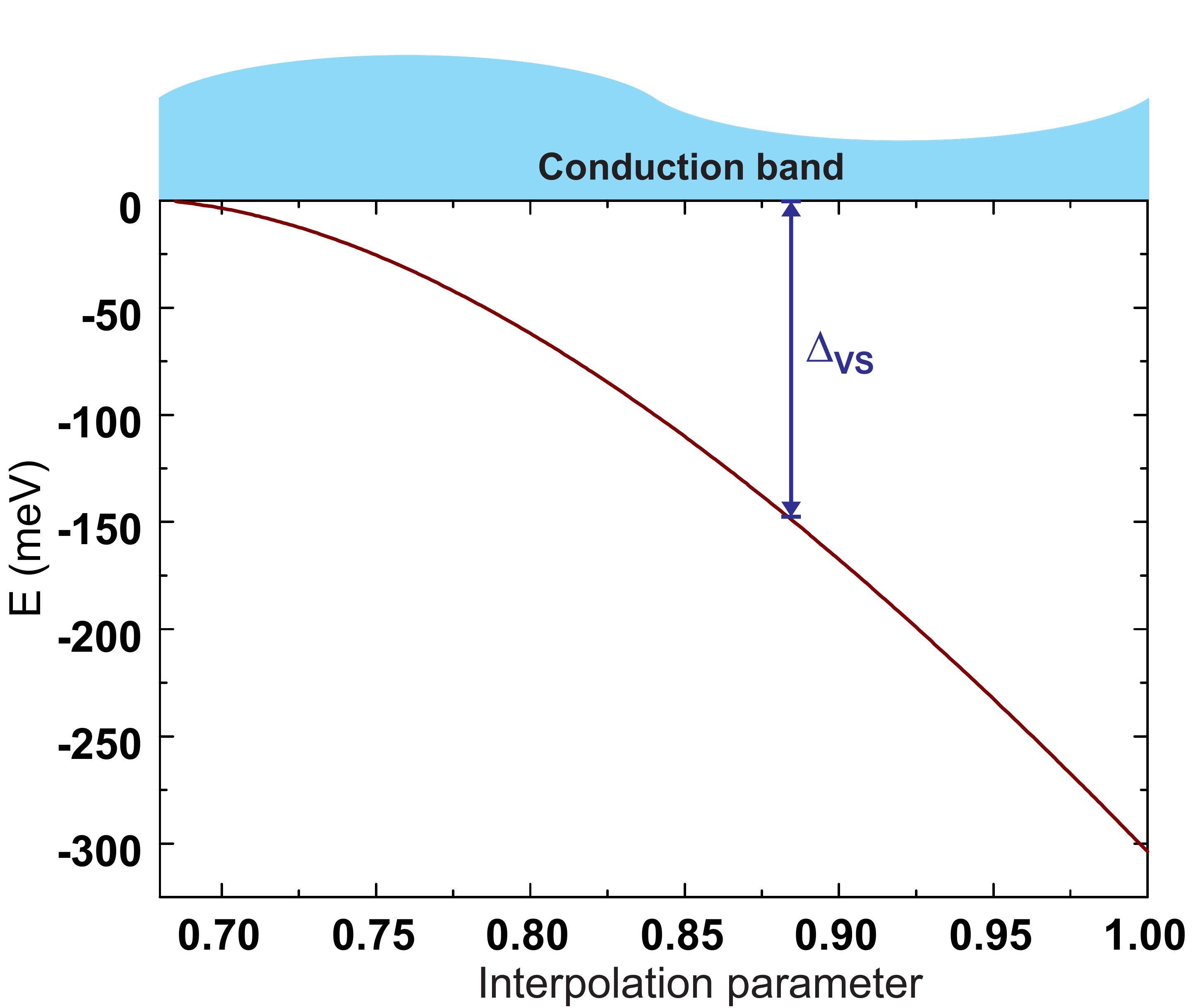}
		\caption{\label{fig:InterfaceEnergy}(Color online) 
Energy eigenvalue of the interface state as  a function of the interpolation parameter $\alpha$.
We identify the valley splitting as the energy gap between the two lowest energy eigenstates, \textit{i. e.}, the band edge and bound state eigenvalue.
}
		\end{figure}

Figures~\ref{fig:LDOS}(c) and \ref{fig:LDOS}(d) present the LDOS at the
junction Si and barrier sites calculated for $\alpha=0.5$ and $1.0$.
The sharp peak below the Si conduction-band edge, marked by an arrow in Fig.~\ref{fig:LDOS}(d), indicates the formation of a TS interface state, while  for $\alpha = 0.5$ the LDOS is strongly modified as compared to the bulk materials in Fig.~\ref{fig:LDOS}(b), but no isolated peak appears. These results suggest the existence of a minimum $\alpha$ needed to form a bound interface state (TS) below the conduction band. In fact, Fig.~\ref{fig:InterfaceEnergy} shows that a TS state appears only for $\alpha\gtrsim 0.7$.
From Fig.~\ref{fig:InterfaceEnergy} it is also clear that the bottom of the conduction band (horizontal line) and the interface state constitute the two lowest
electronic energy levels in this system, i.e., their energy difference gives
the ground-state gap, which we identify here with the valley splitting ($\Delta_{\textrm{VS}}$).
For all $\alpha$, $\Delta_{\textrm{VS}}(\alpha)$ is a single-valued increasing function so that, instead of $\alpha$, the physically accessible quantity $\Delta_{\textrm{VS}}$ is taken as the control variable in terms of which all calculated properties are discussed next. Thus quantities related to regions $I$ and $B$ (see Table I), including $\alpha$, are not expected to affect the main results.

\begin{figure}[ht!]
		\includegraphics[width=\columnwidth]{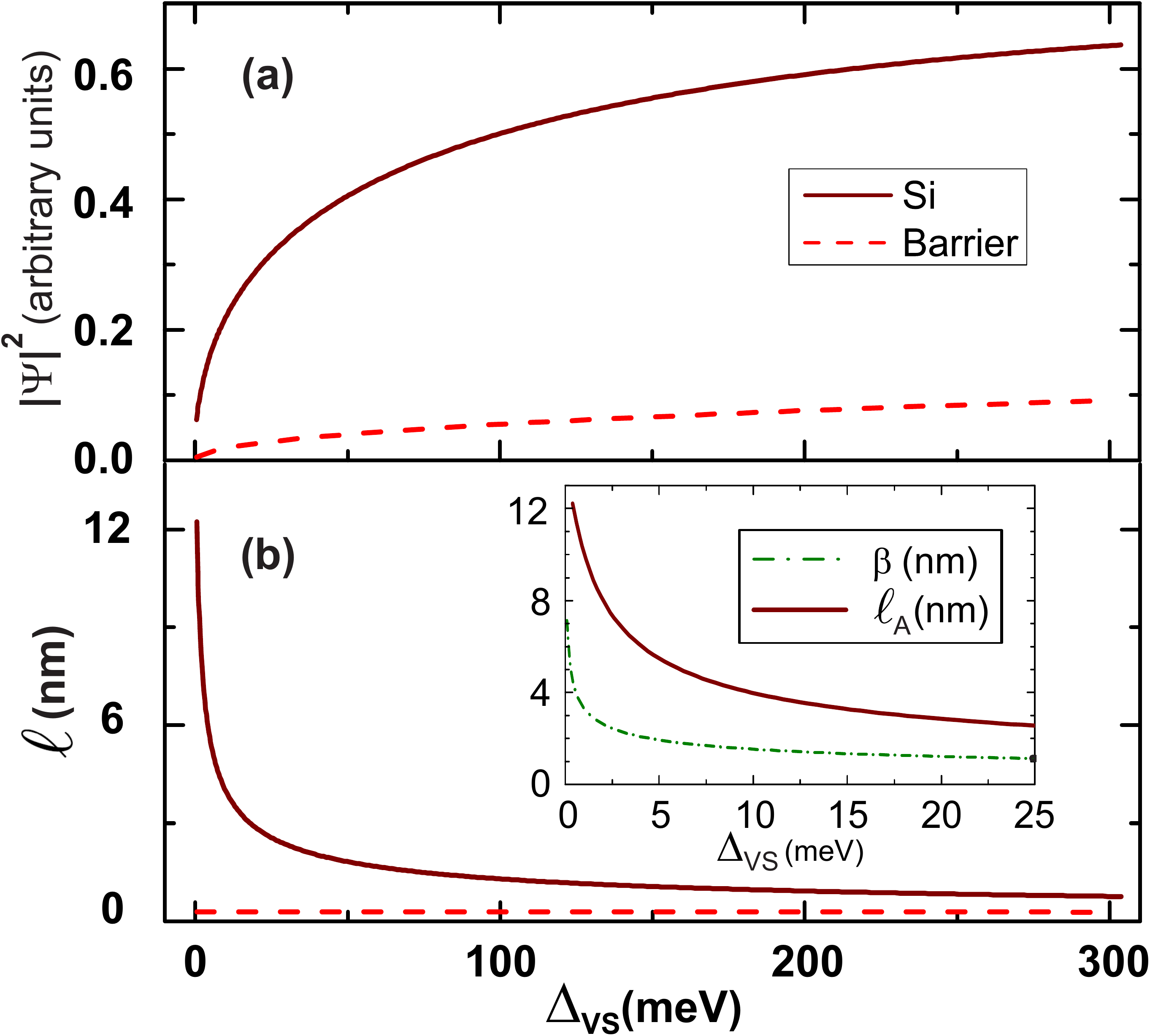}
	   	\caption{\label{fig:AllXEnergy} (Color online) 
Properties of the intrinsic interface state as functions of the valley splitting ($\Delta_{\textrm{VS}}$). (a) Electronic probability density ($|\Psi_s|^2$) at the Si $(s=-1)$ and barrier $(s=1)$ junction sites. (b) Localization length ($\ell$) into Si $(s \leq -1)$ and barrier $(s \geq 1)$ regions.
The inset shows length scales characteristic of the TS and FH states (see Sec. \ref{sec:triangle}) as a function of $\Delta_{\textrm{VS}}$.}
		\end{figure}

		\begin{figure}[h!]
		\includegraphics[width=\columnwidth]{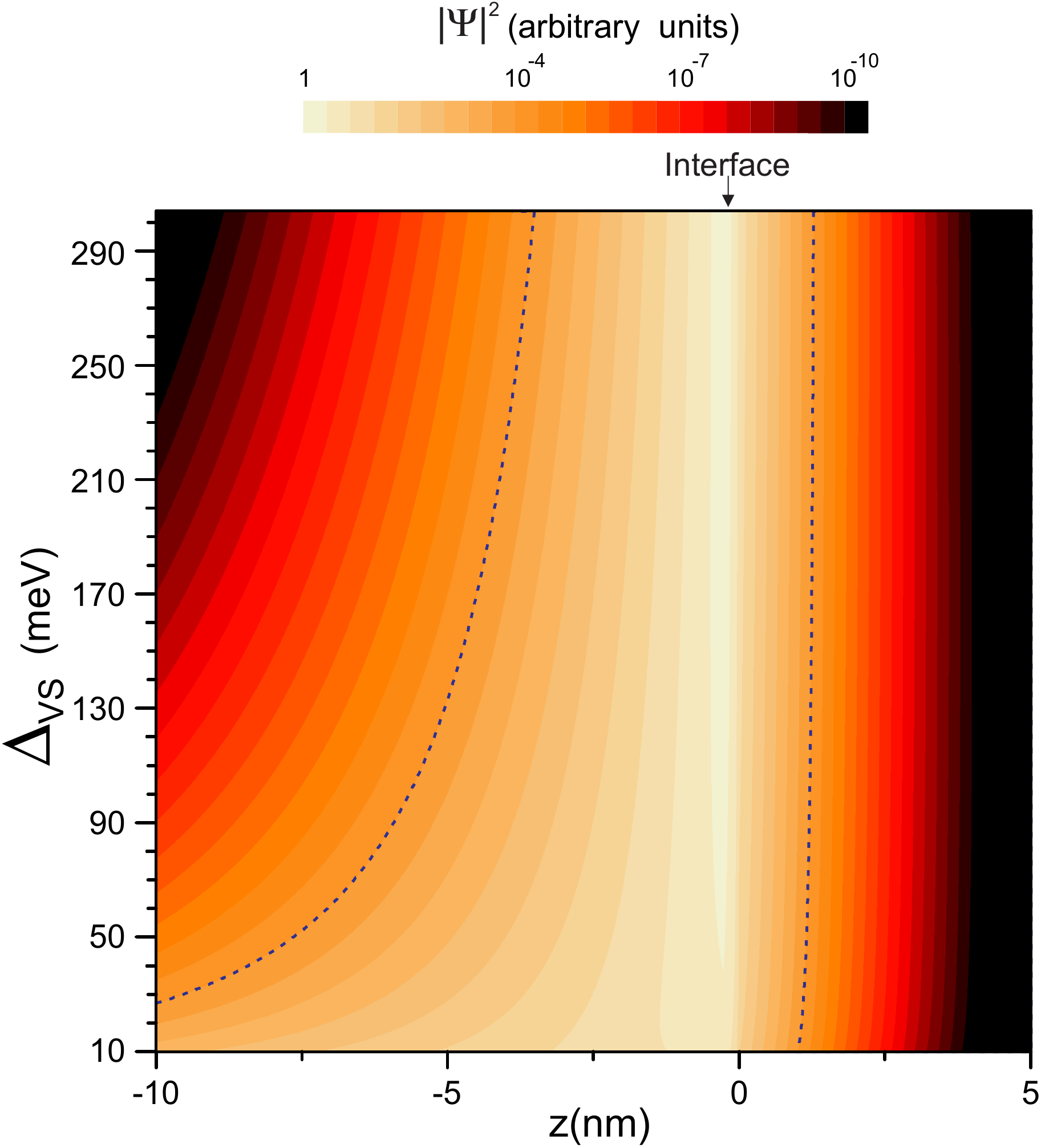}
		\caption{\label{fig:EletronicDensity}(Color online)
 Approximate electronic density of the TS state extracted from the envelope function
asymptotic behavior along the chain as a function of the energy of the ground state ($\Delta_{\textrm{VS}}$), where $z\leq-1$ corresponds to A sites (Si), $z\geq1$ corresponds to B sites (barrier).
Values for the interface sites in Si and barrier were obtained from direct calculation and we assume exponential decay beyond these sites, which should apply to $|s|\gg1$.}
		\end{figure}

Figure~\ref{fig:AllXEnergy}(a) shows the local density of states at $(s=\pm 1)$, or equivalently the
electronic probability density $|\Psi_{\pm 1}(E=-\Delta_{\textrm{VS}})|^2$ of the TS
state at the  interface sites as a function of the valley splitting.
Operationally, these are the weight of the pole of $G_{1 1}$  and $G_{-1-1}$ at the energy of the interface state $(E=-\Delta_{\textrm{VS}})$.

The asymptotic behavior of localized states away from the perturbation site(s) is  characterized by an envelope exponential decay within a distance $\lambda$,
	\begin{equation}
	\label{eq:Asymp}
	lim_{s \rightarrow \pm\infty}| \Psi(s) | \propto \exp ({-{|s|}/{\lambda}})
	\end{equation}
assuming that the perturbation creating the localized state is symmetric.
For nearest neighbors tight-binding alone, it was shown\cite{Herbert1971,Thouless1972,Robbins1985} that
	\begin{equation}
	\label{eq:Localization}
\lambda^{-1}=-\lim_{n \rightarrow \infty}2^{-n}\ln |t^{(n)}|,
	\end{equation} where $t^{(n)}$ is the nearest-neighbors hopping parameter
after $n$ decimation cycles (see Appendix \ref{app:decimation}).

The envelope function of a TS state is expected to be asymmetric with respect to the junction so Eq. (\ref{eq:Asymp}) splits into two localization lengths: $ \ell_{\rm A}$ into the Si (s $\rightarrow -\infty$) and $ \ell_{\rm B}$
into the oxide side (s $\rightarrow +\infty$).

We expect that expressions analogous to (\ref{eq:Localization}) apply in the present case (this assumption is verified numerically \textit{a posteriori}). Since for the second-neighbors model the intermediate decimated chain configurations acquire nonzero hopping parameters up to fourth nearest neighbors (see Appendix \ref{app:decimation}), a possible generalization of Eq.~(\ref{eq:Localization}) would be written as
\begin{equation}
	\label{eq:GenLocalization}
            \ell^{-1}_{k=2,3,4}({\rm K})=-\lim_{n \rightarrow \infty}{C_k^{(n)}}\, {\ln |t^{(n)}_{k}({\rm K})|},~{\rm K=A,B},
        \end{equation}
where $\lim_{n \rightarrow \infty}{C_k^{(n)}}= 2^{-(n+1)}$ and $t^{(n)}_{k}({\rm K})$ for $k=$ 2, 3, 4 are the renormalized $k$th-neighbors hopping parameters after the $n$th decimation cycle at either $s\leq-1~(A)$ or $s\geq1~(B)$. 
In fact,  the limits for all ranges $k$ converge to \textit{the same} $\ell_A$~or~$\ell_B$,  indicating the consistency of our assumption. 
Figure \ref{fig:AllXEnergy}(b) gives $\ell_A$ (solid line) and $\ell_B$ (dashed line) versus $\Delta_{\textrm{VS}}$.
The expected behavior is obtained in Fig.~\ref{fig:AllXEnergy}(b), namely, a TS state tends to delocalize when approaching the Si band edge due to band-localized state hybridization.
Figure~\ref{fig:EletronicDensity} summarizes the behavior of the interface states  localization lengths.
We take the exponential decay in the probability density---consistent with $\ell$ given in Fig.~\ref{fig:AllXEnergy}(b) for each material---to apply for all sites beyond the respective junction sites $s=\pm 1$. 
We show in Fig.~\ref{fig:EletronicDensity} that the electronic probability peak at the interface rises as the TS state becomes deeper in energy, reducing the penetration into the Si slab. 
Penetration is qualitatively indicated by the color stripes. 
As a general (arbitrary) guide, the dashed line follows the point where the density is reduced from the interface peak by 4 orders of magnitude. 
The penetration into the barrier is negligible with respect to the Si side, and not very sensitive to the TS energy. 
The characteristic oscillatory behavior of localized states in Si---due to the interference of $k_z$ and $k_{-z}$ valleys---is not reproduced here.~\cite{saraiva2010}

\section{\label{sec:triangle}Electric Field Bound States}

A more familiar interface-bound electron state is formed when an applied electric field ($F$) pushes the electron towards an interface.
A number of studies on this problem based on tight-binding treatments are available in the literature.~\cite{Grosso1996,Boykin2004,Boykin2004PRB,saraiva2010}
In the presence of an electric field, solving  a tight-binding model based on the renormalization method is cumbersome and not specially insightful. 
In particular, localization properties are not easily extracted as in Sec. \ref{sec:InterfaceState} [see Eq. (\ref{eq:Localization})].

Overall insight of interface states bound by an electric field, including localization trends, may be obtained within a simplified description proposed by Fang and Howard (FH)~\cite{fang1966}, which we briefly review. 
This approach is based on the effective mass envelope function, providing analytic expressions within a single valley approximation.
In FH's model the barrier region ($z>0$) is assumed to be impenetrable, and the effective Hamiltonian for $z<0$, with $m_z$ as the longitudinal effective electron mass in Si, $e$ the absolute electron charge, and $\varepsilon_{\rm A}$ the Si relative permittivity, is written as
\begin{equation}
\label{eq:HamiltonianTriangularPotential}
	H_{F\!H}=-\frac{\hbar^2}{2m_z}\frac{\partial^2}{\partial z^2}+e\frac{F}{\varepsilon_{\rm A}}|z|.
\end{equation}
The FH variational envelope for the  ground state~\cite{fang1966} is assumed into have the form
\begin{equation}\label{eq:tentative_WaveFunction}
  \Psi_{F\!H}(z<0) =
      \sqrt{{4}/{\beta^3}} |z| e^{-|z|/{\beta}} 
\end{equation}
and $\Psi_{F\!H}(z>0)=0$. The parameter $\beta$ characterizes the wavefunction extent into the Si slab and is obtained from energy minimization,
       \begin{equation}\label{eq:b}
            \beta=\left( \frac{2 \hbar^2 \varepsilon_{\rm A}}{3 m_z e F} \right)^{1/3}.
        \end{equation}
The lengths $\beta$ and $\ell_A$ are given in the inset of Fig.~\ref{fig:AllXEnergy}(b). Further details and impact of these different lengths are discussed in the next section.

We have also investigated electric field bound states within tight binding using supercells direct diagonalization (see Appendix~\ref{app:supercell}). In what follows we refer to electric field bound states as FH states, even if they are obtained within tight-binding.

      \section{\label{sec:field} Field Control of the Splitting}

The valley splittings of the field bound FH states and the spontaneously bound TS states have distinct origins, but both depend on the external electric field. The difference between the mechanisms is a subtlety that results in immensely different splitting-to-field rates.

Traditional conduction-band FH states have the same overall shape for the envelope function for each valley composition, making it possible to study it under the effective-mass approximation~\cite{Saraiva2009}. This means that the electronic density is very similar and the electric field does not couple directly to the valley degree of freedom. Instead, the role of the field is indirect, squeezing the electron against the barrier material.

On the other hand, the effect of electric fields in spontaneously localized interface TS states is trivial. 
The average position of the electronic charge distribution for the TS and FH states is different, so that the electric field detunes them.
The TS state is roughly an exponential with an average position $\langle z \rangle = \ell_{\rm A}/2$, while a regular conduction band FH state is located at $\langle z \rangle = 3 \beta/2$.

\begin{figure}[h]
		\includegraphics[width=\columnwidth]{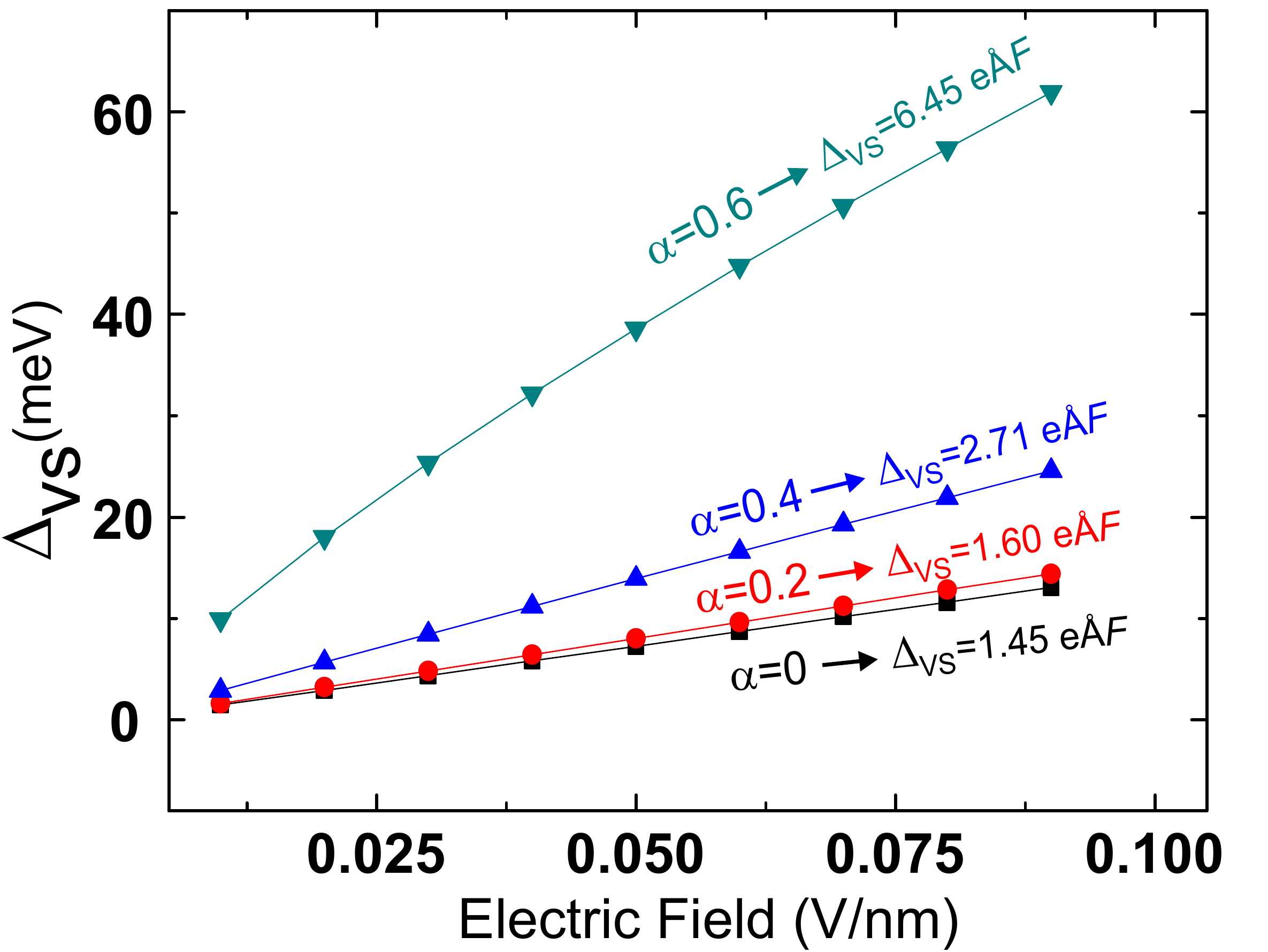}
		\caption{\label{fig:alpha-dependence}(Color online) Electric field dependence of the valley splitting for different values of $\alpha$, as indicated. Linear fits for each $\alpha$ are also given. Supercell:  67.9 nm Si layer adjacent to a 6,8 nm oxide layer.}
		\end{figure}

The valley splitting sensitivity to the charge distribution is illustrated in Fig.~\ref{fig:alpha-dependence}, where we show the $\Delta_{\textrm{VS}}$ dependence on the external field obtained within the same tight-binding model, defined in Eq.~\ref{eq:Hamiltonian}, via a supercell methodology (described in Appendix \ref{app:supercell}). Varying the value of $\alpha$ we identify a crossover of the splitting from FH behavior  (low $\alpha$) to the typical enhancement of $\Delta_{\textrm{VS}}$ characteristic of important hybridization of the interface state with band states  (higher $\alpha$). Comparison of the $\Delta_{\textrm{VS}}$ sensitivity in $F$ as $\alpha$ increases in the range  $\{0.2 \to 0.4\}$ to the range $\{0.4 \to 0.6\}$ shows that the TS regime is approaching at $\alpha = 0.6$. In fact for $\alpha = 0.8$ the data would extrapolate the plotted range.

These results are more clearly understood by examining the two lower eigenstates for FH and TS states, obtained as described in Appendix~\ref{app:supercell}, and given in Fig.~\ref{fig:ts-vs-fh}. The FH state---illustrated taking  $\alpha = 0$---presents the usual behavior of electrons bound by the triangular electrostatic potential, with a splitting consistent with first order perturbation theory.\cite{Saraiva2009}
The valley splitting, identified here as the energy difference between these states and shown in the inset, increases linearly with the field, remaining  of the order of a few meV.

We take $\alpha=0.5$ to model near band-edge TS states. Although at $F=0$ this value of $\alpha$ is below the minimum required to sustain an interface state, a TS state splits from the band into the gap at very small electric fields. The lower energy level behavior (black squares) for the TS state, shown in Fig.~\ref{fig:ts-vs-fh}(b), is quantitatively very different from the behavior in the FH state [Fig.~\ref{fig:ts-vs-fh}(a)]. The first excited state values (red circles), however, are closer to each other: this is consistent with our interpretation of the origin of the enhanced valley splitting.

		\begin{figure}[h]
		\includegraphics[width=\columnwidth]{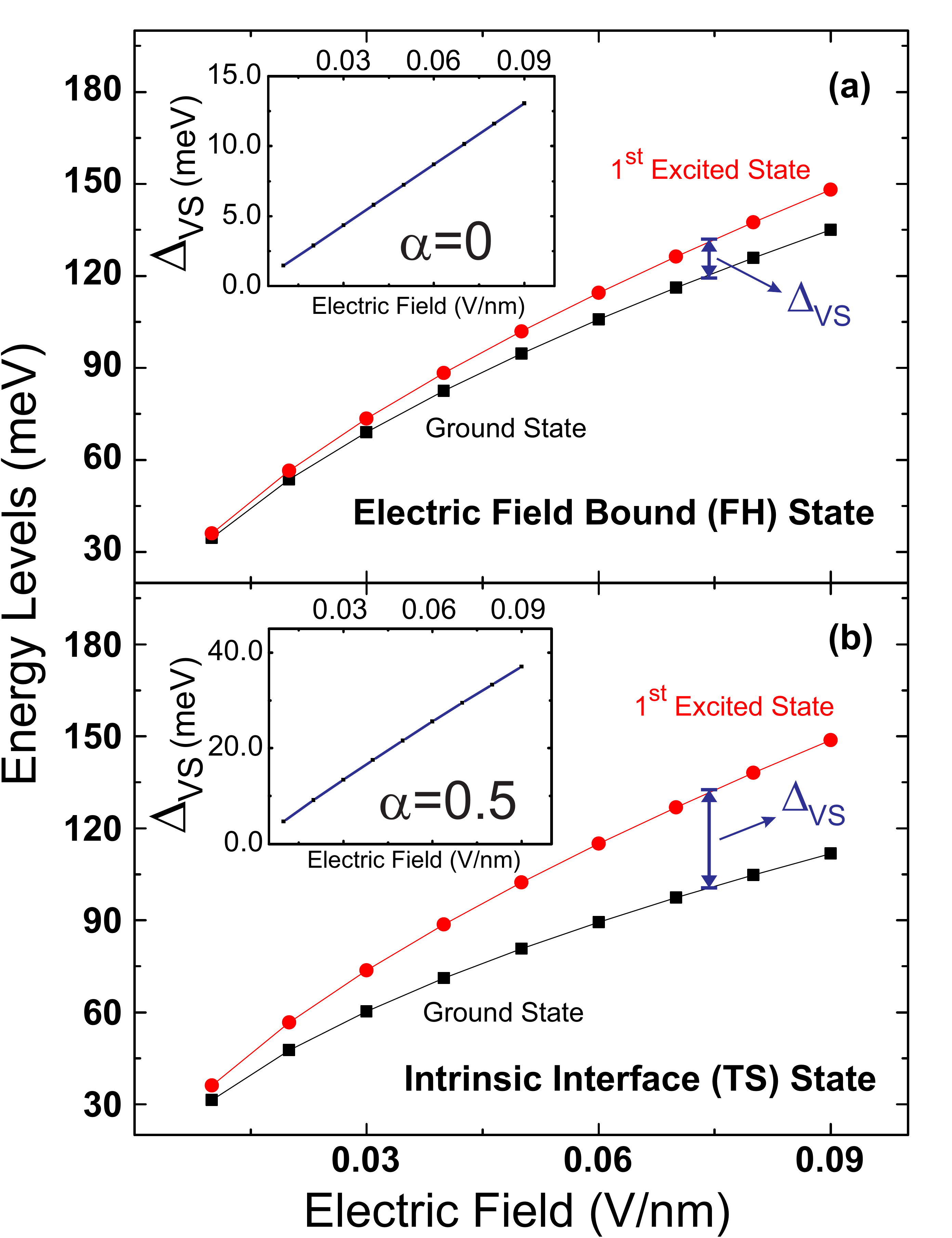}
		\caption{\label{fig:ts-vs-fh}(Color online) (a) Electric field dependence of ground and first excited energy levels calculated within tight binding for an electric field bound state ($\alpha=$0). The inset shows the energy difference between theses states, i.e., the $\Delta_{\textrm{VS}}$. (b) Same as (a) for an intrinsic interface state ($\alpha=$0.5). Supercell:  67.9 nm Si layer adjacent to a 9.5 nm oxide layer.}
		\end{figure}

The insets in Fig.~\ref{fig:ts-vs-fh} show that the fundamental difference between the origin of FH and TS states leads to distinct responses to applied electric fields. This distinction causes the TS state to present improved tunability of the valley splitting, covering a wider range of $\Delta_{\textrm{VS}}$.

\section{\label{sec:Capacitance}Capacitance and interface states }

We investigate in this section to what extent capacitance measurements may distinguish TS from FH interface states, given the distinct real space electronic charge
distributions of these states.

For both TS and FH interface states, the penetration into the barrier side is negligible, therefore we do not take the contribution of the fast decaying charge density tail into the barrier. 
Also, the oscillatory behavior mentioned in Sec.~\ref{sec:InterfaceState} is expected to make comparable contributions to both TS and FH localized states and should not alter significantly the comparison performed here.

\begin{figure}[ht!]
		\includegraphics[width=0.99\columnwidth]{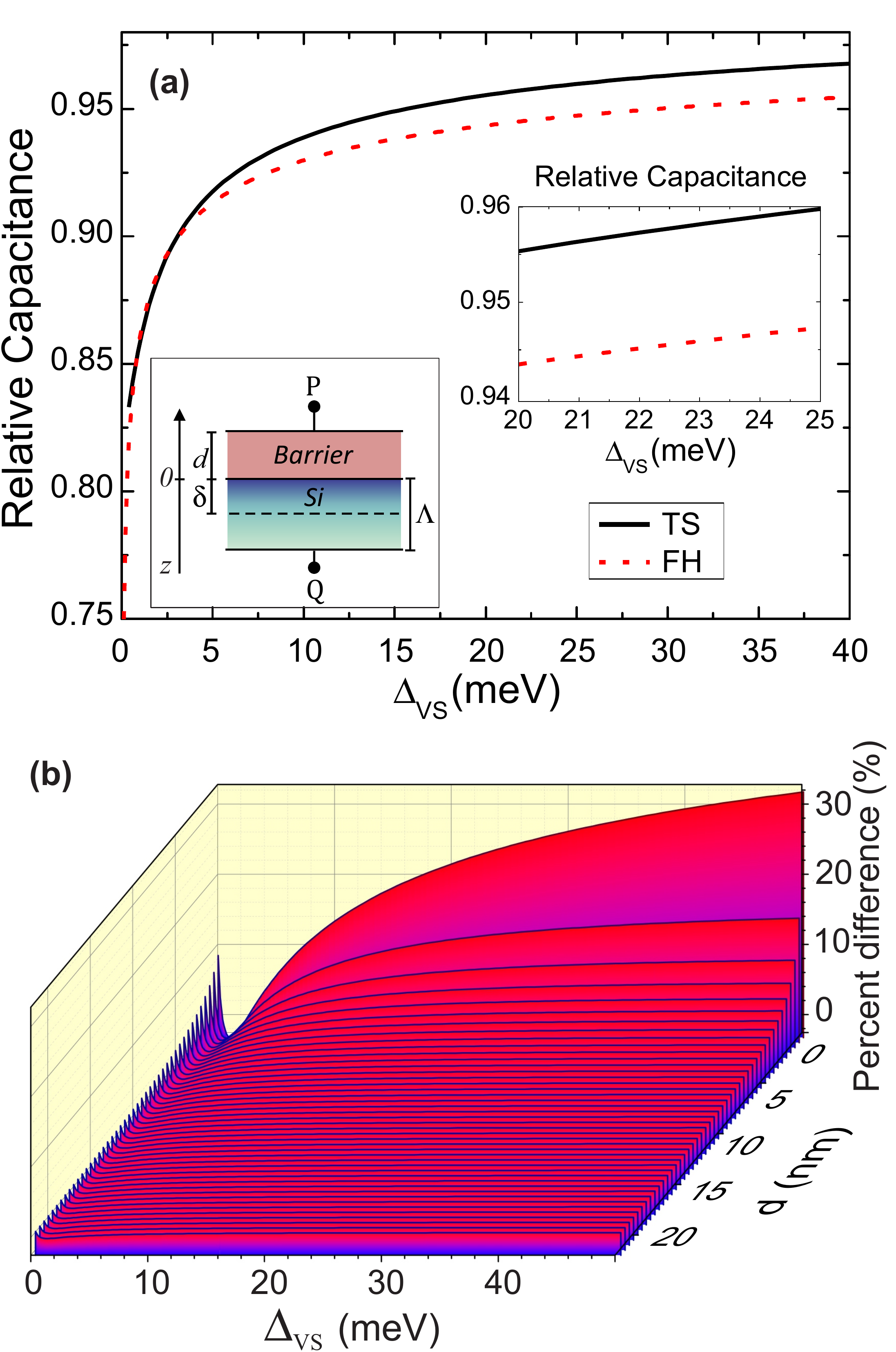}
		\caption{\label{fig:CapacitanceXDVO}(Color online) 
(a) Relative capacitance for  TS and FH interface states as a function of the valley splitting. 
The lower-left inset shows a  schematic representation of the capacitor structure discussed here, $\delta$ is the effective thickness of the equivalent capacitor. 
The length $\Lambda$ is large enough to include the total charge bound near the interface.
(b) Percent difference between capacitances in the presence of a TS or FH state as a function of the valley splitting $\Delta_{\textrm{VS}}$ and the barrier thickness $d$.}
		\end{figure}

The capacitance between the gate $P$ and the electron gas is considered under the geometry schematically shown in the lower-left inset of
Fig.~\ref{fig:CapacitanceXDVO}(a), taken to be parallel plates capacitors.
The exponential charge penetration of a TS state into the Si layer is
        \begin{equation}	\label{eq:TSCharge}
        \rho_{\textrm{TS}}(z)=\rho_0 e^{-{2|z|}/{\ell_{\rm A}}}.
        \end{equation}
The dielectric barrier of thickness $d$ gives a capacitance
   	\begin{equation}
	\label{eq:InterfaceCapacitance}
        C_{\textrm{TS}} (\ell)=\frac{{\cal A}\varepsilon_{0}\varepsilon_{\rm A}\varepsilon_{\rm B}}{\varepsilon_{\rm A} d+(1/2)\varepsilon_{\rm B}\ell_{\rm A}},
        	\end{equation}
where $\varepsilon_{\rm A}$ ($\varepsilon_{\rm B}$) is the relative permittivity of the silicon (barrier) and  $\cal A$ is the device's transverse area.

If instead the gas forms by the usual field binding, the charge density will be derived from the FH wave function [see Eq.~\ref{eq:tentative_WaveFunction})],
\begin{equation}\label{FHdensity}
        \rho_{\textrm{FH}}(z)= \left({4}/{\beta^3}\right)|z|^2 \rho_0 e^{-2{|z|}/{\beta}}.
\end{equation}
In this case, the capacitance reads
	\begin{equation}
	\label{eq:FHCapacitance}
            C_{\textrm{FH}}=\frac{{\cal A}\varepsilon_{\rm 0}\varepsilon_{\rm A} \varepsilon_{\rm B}}{\varepsilon_{\rm A} d + ({3}/{2}) \varepsilon_{\rm B} \beta }.
        	\end{equation}

 The effective thickness $\delta$ of the electron charge distribution for the TS (FH) state is ${\ell_A}/{2}$ (${3\beta}/{2}$) [$\delta$ is schematically indicated in the lower-left inset in Fig.~\ref{fig:CapacitanceXDVO}(a), and is assumed to be much smaller than the Si slab].

The capacitances share the same overall form---the contribution of the electron penetration into the semiconductor may be regarded as a parasitic capacitance associated in series with the otherwise ideally thin capacitor $C_0={{\cal A} \varepsilon_{\rm 0}\varepsilon_{\rm B}}/{d}$. Therefore, capacitance measurements are only capable of distinguishing these states if the base capacitance $C_0$ due to the SiO$_2$ barrier is comparable ($\gtrsim$) to the parasitic capacitances.

Since the detailed electric field inside a semiconductor heterostructure may be hard to obtain (especially for doped samples), we calculate all quantities as a function of $\Delta_{\textrm{VS}}$. For the interface (TS) state, the splitting shown is the zero-field spontaneous splitting, while for the external field bound (FH) state we assume a field high enough to give each  $\Delta_{\textrm{VS}}$. Theory~\cite{Saraiva2011} indicates that this mechanism leads to $\Delta_{\textrm{VS}}=0.548\, e$\,\AA$F$, which is in fair agreement with recent experiments~\cite{yang_2013}.

The lengths $\ell_{\rm A}$ and $\beta$ are presented in the inset of Fig.~\ref{fig:AllXEnergy}(b). Even though $\ell_{\rm A} > \beta$, the parasitic capacitance expressions for TS and FH states depend on $\ell_{\rm A}/2$ and $3\beta/2$, respectively. Taking this into account, there is an inversion at $\Delta_{\textrm{VS}} \gtrsim 5$ meV, resulting in the capacitance relation $C_{\textrm{TS}} > C_{\textrm{FH}}$.

Figure~\ref{fig:CapacitanceXDVO}(a) shows the capacitances (in $C_0$ units) as a function of $ \Delta_{\textrm{VS}}$. We chose here $d=10$ nm, leading to a small capacitance contribution from the oxide (further reduced by the small oxide relative permittivity $\varepsilon_{\rm B}=3.9$, compared to Si, $\varepsilon_{\rm A}=11.9$).
Figure~\ref{fig:CapacitanceXDVO}(b) gives the percent difference $D_p=(C_{\textrm{TS}}-C_{\textrm{FH}})/C_{\textrm{TS}}$ as a function of the oxide thicknesses $d$ and the splitting $ \Delta_{\textrm{VS}}$.

We first note that $C_{\textrm{TS}} > C_{\textrm{FH}}$ except for a small range in the low-$\Delta_{\textrm{VS}}$ limit, which appears in the figure as a dip next to the peak at $\Delta_{\textrm{VS}}=0$. For $\Delta_{\textrm{VS}} = 23$ meV, $D_p= 1.7\%$ [as may be obtained from the zoomed region in Fig. \ref{fig:CapacitanceXDVO}(a)]. It is clear that smaller values of $d$  increase the distinguishability between the two kinds of interface states.

Reliable identification of the presence of intrinsic interface states (TS) for a particular Si/SiO$_2$ junction may be possible through a comparison of different samples and/or different interfaces on the same sample as in Refs.~[\onlinecite{niida2013,takashina2006,takashina2004,ouisse1998}]. General trends towards this identification obtained here for the interface TS states contribution to capacitance as compared to the more usual FH state may be summarized as follows: (i) at low voltages, $C_{T\!S}$ should be less sensitive to voltage variations than $C_{F\!H}$ and remain finite in the limit of very small voltages; (ii) At the conditions leading to the same splitting $\Delta_{\textrm{VS}}$, the interface TS state will have a larger capacitance than the field bound FH state -- an effect enhanced for thin oxide barriers.

		\section{Discussions and Conclusions}
We have presented a comprehensive study of properties and effects of interface states in a Si/SiO$_2$ interface.
The adopted model,  based on a simple one-dimensional tight-binding description of the system, is solved exactly within a Green's-function renormalization formalism through an original decimation sequence, also presented. From the converged fixed point of the decimation procedure, the local density of states at the interface sites is obtained, as well as the energies and conditions for the formation of an intrinsic interface state. From the Green's-function off-diagonal matrix elements, the rate of approach to the fixed point gives the localization lengths characterizing the exponential decay of the interface state into each of the materials in the heterojunction.

For comparison, we have also considered electric field bound states within the FH effective-mass approximation---inferring localization trends and the overall charge distribution used in calculations of the capacitance. 
The distinct charge distributions of these two types of states, with a peak (TS) or node (FH) at the interface, are explored in terms of differences in capacitance measured for a voltage bias applied at opposite sides of the interface.
We find that in the range of high-$\Delta_{\textrm{VS}}$, the calculated $C_{\textrm{TS}}$ is always larger than $C_{\textrm{FH}}$, and that the distinction between them is enhanced for narrower barrier widths. Such differences are expected to be observable and useful to verify experimentally whether intrinsic interface states (TS) are present and probably responsible for observed  high-$\Delta_{\textrm{VS}}$ values of some junctions when different samples and interfaces are compared.

 Field control over the ground and first excited levels of FH and TS states were also obtained in a supercell tight-binding approach. The TS state is clearly shown to improve the tunability of the valley splitting over a wider range.

\section{Acknowledgements}

We are indebted to Kei Takashina, Mark Friesen, and Rodrigo Capaz for the many discussions.
This work is part of the
Brazilian National Institute for Science and Technology
on Quantum Information. The authors also acknowledge
partial support from FAPERJ, CNPq, and CAPES.
\vfill

\bibliography{Bibliography}

\bigskip\bigskip
\appendix
\section{\label{app:decimation} Dimer decimation method}

\begin{figure}[h!]
		\includegraphics[width=\columnwidth]{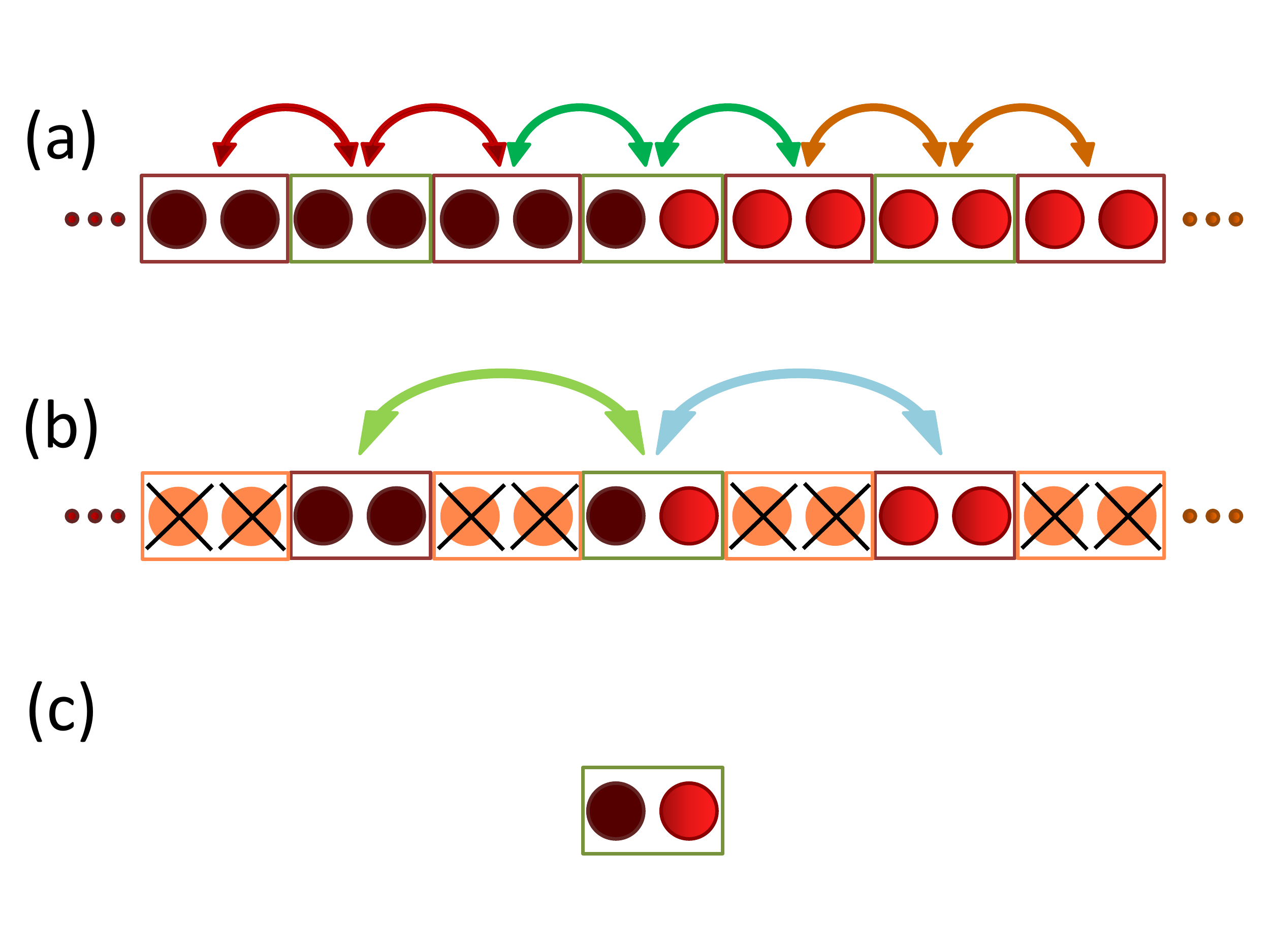}
		\caption{\label{fig:Decimation}(Color online) 
(a) Original chain (prior to decimation), considering the nearest-neighbors dimers coupling. 
(b) Intercalated dimers are decimated each cycle, and renormalized hoppings connect the next-nearest dimers. 
(c) Final product of the decimation process, i.e., a single dimer with negligible coupling with nearest dimers. After the process of decimation, the original system is represented by two effective atoms with complex tight-binding parameters. }
		\end{figure}

We start with Dyson's equation for all sites,
\begin{equation}
\label{eq:Dyson}
Z~G_{ij}-\sideset{}{'}\sum_{l}H_{il}~G_{lj}=\delta_{ij},
\end{equation}
representing each dimer as a pair $(i,j)$, and discarding the label $0$, for symmetry of the decimation procedure, in the first renormalization cycle, all information  involving alternate dimers---$(2,3)$ $(6,7)$ $(10,11)$ ... $(2+4^N,3+4^N)$ to the right and $(-2,-3)$ $(-6,-7)$ $(-10, -11)$ ... $ (-2 -4^N,-3 -4^N)$ to the left is decimated [see Figs.~\ref{fig:Decimation}(a) and (b)],   i.e., projected  into those for the ``surviving" sites, resulting in a problem of closely bound pairs separated by a distance that eventually increases exponentially with the number of decimation cycles.

For example, after the first cycle, $(-1,1)$ and $(4,5)$ are nearest-neighbor dimers, with hoppings defined as in Fig.~\ref{fig:Dimers}. Always keeping information on
$(-1,1)$, the second cycle consists of projecting out the dimers' nearest
neighbors to it, namely $(4,5)$, $(-4,-5)$  , so that  $(8,9)$  becomes  a nearest
neighbor dimer to $(-1,1)$ at the end of the second cycle.

The possible hopping elements in the original ($n-$ times decimated)  chain
are given  in Tables II and III, where the notation is self-explanatory. In
Table III, the fixed point ($n\to\infty$)  is 0 for all $t_2$, $t_3$ and $t_4$,
while $t_1$ converges to $t_1  (A)$, $t_1 (B)$, or $t_1 (I)$ according to
the constituents of the dimers.

The fixed point is an isolated dimer, (-1,1) [see Fig.~\ref{fig:Decimation} (c)],   obtained when all $t_i$ for $i=2,3,4$ are close to 0 (within a convergence criterion that we take as $u*10^{-3}$, with $u$ introduced in the definition $\hat G(Z)=\widehat{G}(E+iu)$ in Sec. \ref{sec:Formalism}. All needed matrix
elements $G_{-1 -1}, G_{1 1}$, and $G_{1 -1}=G_{-1 1}$ are easily identified at the fixed point.

\begin{table*} [ht!]
 \caption{\label{tab:Dimers_combinations_0}
  Hopping elements of the original Hamiltonian $(n=0)$. Here $A$ indicates Si sites, $B$ indicates barrier sites and $I$ indicates interface sites. See Eq.~\ref{eq:Hamiltonian} for the notation, where the superscript (0) is added here and site labels are defined in Fig.~\ref{fig:Dimers}.
   }
	\begin{ruledtabular}
		\begin{tabular}{c c c c c c c c}
	 & & & & & & & \\
	&$\kappa~\lambda~\mu~\nu$ & $t_{1}^{(0)}\left[\kappa \xrightarrow{} \lambda\right]$ &  $t_{1}^{(0)}\left[\mu \xrightarrow{} \nu\right]$ &  $t_{2}^{(0)}\left[\lambda \xrightarrow{} \mu \right]$ &  $t_{3}^{(0)}\left[\kappa \xrightarrow{\lambda} \mu \right]$&  $t_{3}^{(0)}\left[\lambda \xrightarrow{\mu} \nu\right]$ & $t_{4}^{(0)}\left[\kappa \xrightarrow{\lambda \mu} \nu\right]$ \\

	 & & & & & & & \\
	 & & & & & & & \\
	 Region $A$ & $A$ $A$ $A$ $A$ & $t_{1}(A)$ &$t_{1}(A)$ &$t_{1}(A)$ & $t_{2}(A)$&$t_{2}(A)$ &0 \\
	 & & & & & & & \\
	 & & & & & & & \\
	 &$A$ $A$ $A$ $B$ & $t_{1}(A)$ & $t_{1}(I)$ & $t_{1}(A)$ & $t_{2}(A)$ &  $t_{2}(I)$ & 0 \\
	Interface & & & & & & & \\
	 &$A$ $B$ $B$ $B$ & $t_{1}(I)$ & $t_{1}(B)$ & $t_{1}(B)$ & $t_{2}(I)$ & $t_{2}(B)$ & 0  \\
	 & & & & & & & \\
	 & & & & & & & \\
	Region $B$&$B$ $B$ $B$ $B$ & $t_{1}(B)$ & $t_{1}$(B) & $t_{1}(B)$ & $t_{2}(B)$ & $t_{2}(B)$ & 0 \\
	 & & & & & & & \\
		\end{tabular}
	\end{ruledtabular}
\end{table*}

\begin{table*} [ht!]
 \caption{\label{tab:Dimers_combinations_n} Same as Table \ref{tab:Dimers_combinations_0} for the $n$-times decimated chain. All sites in neighboring dimers are coupled (see Fig.~\ref{fig:Dimers} for the notation). }
	\begin{ruledtabular}
		\begin{tabular}{c c c c c c c c}
	 & & & & & & & \\
	&$\kappa~\lambda~\mu~\nu$ & $t_{1}^{(n)}\left[\kappa \xrightarrow{} \lambda\right]$ &  $t_{1}^{(n)}\left[\mu \xrightarrow{} \nu\right]$ &  $t_{2}^{(n)}\left[\lambda \xrightarrow{} \mu \right]$ &  $t_{3}^{(n)}\left[\kappa \xrightarrow{\lambda} \mu \right]$&  $t_{3}^{(n)}\left[\lambda \xrightarrow{\mu} \nu\right]$ & $t_{4}^{(n)}\left[\kappa \xrightarrow{\lambda \mu} \nu\right]$ \\

	 & & & & & & & \\  \hline
	 & & & & & & & \\
	 Region $A$ & $A$ $A$ $A$ $A$ & $t_{1}^{(n)}\left[A \xrightarrow{} A\right]$ &  $t_{1}^{(n)}\left[A \xrightarrow{} A\right]$ &  $t_{2}^{(n)}\left[A \xrightarrow{} A\right]$ &  $t_{3}^{(n)}\left[A \xrightarrow{A} A\right]$&  $t_{3}^{(n)}\left[A \xrightarrow{A} A\right]$ & $t_{4}^{(n)}\left[A \xrightarrow{AA} A\right]$ \\
	 & & & & & & & \\
	 & & & & & & & \\
	 &$A$ $A$ $A$ $B$ & $t_{1}^{(n)}\left[A \xrightarrow{} A\right]$ &  $t_{1}^{(n)}\left[A \xrightarrow{} B\right]$ &  $t_{2}^{(n)}\left[A \xrightarrow{} A\right]$ &  $t_{3}^{(n)}\left[A \xrightarrow{A} A\right]$&  $t_{3}^{(n)}\left[A \xrightarrow{A} B\right]$ & $t_{4}^{(n)}\left[A \xrightarrow{AA} B\right]$ \\
	Interface & & & & & & & \\
	 &$A$ $B$ $B$ $B$ & $t_{1}^{(n)}\left[A \xrightarrow{} B\right]$ &  $t_{1}^{(n)}\left[B \xrightarrow{} B\right]$ &  $t_{2}^{(n)}\left[B \xrightarrow{} B\right]$ &  $t_{3}^{(n)}\left[A \xrightarrow{B} B\right]$&  $t_{3}^{(n)}\left[B \xrightarrow{B} B\right]$ & $t_{4}^{(n)}\left[A \xrightarrow{BB} B\right]$ \\
	 & & & & & & & \\
	 & & & & & & & \\
	Region $B$ & $B$ $B$ $B$ $B$ & $t_{1}^{(n)}\left[B \xrightarrow{} B\right]$ &  $t_{1}^{(n)}\left[B \xrightarrow{} B\right]$ &  $t_{2}^{(n)}\left[B \xrightarrow{} B\right]$ &  $t_{3}^{(n)}\left[B \xrightarrow{B} B\right]$&  $t_{3}^{(n)}\left[B \xrightarrow{B} B\right]$ & $t_{4}^{(n)}\left[B \xrightarrow{BB} B\right]$ \\
	 & & & & & & & \\
		\end{tabular}
	\end{ruledtabular}
\end{table*}

\section{\label{app:supercell} Supercell tight-binding treatment for interface states
under applied electric fields}

In the case of our one-dimensional model under an applied field, the problem is treated within tight binding through a supercell approach. The Hamiltonian in Eq.~\ref{eq:Hamiltonian} is truncated by taking a finite sum in $s$, restricted to a supercell with periodic boundary conditions, which is diagonalized numerically. 
Large enough supercells are used to eliminate  spurious interactions with the periodic images created by these boundary conditions. The same tight-binding parameters described in Sec.~\ref{sec:Formalism} are used, which incorporate correctly the following gap properties: nature, value and relative offset. Data presented in Figs. \ref{fig:alpha-dependence} and \ref{fig:ts-vs-fh} were obtained within this treatment.

Given that the ground-state energy $E_{\textrm{FH}}$ obtained from the optimized variational parameter $\beta$ in Eq.~\ref{eq:b} is proportional to $F^{2/3}$, we explore this relation to probe the consistency of our tight-binding calculations with respect to the FH effective-mass results for the electric field dependence of the ground-state energy. The test here consists in a comparison, the same presented in Ref.~[\onlinecite{Grosso1996}] for a tight-binding three-dimensional model, of the ratio $R=E_{\textrm{FH}}(2 F)/E_{\textrm{FH}}(F)$, which equals $1.59$ in FH. It was found to be equal to $1.4$ at $F=0.04$ V/nm in Ref.~[\onlinecite{Grosso1996}]. We get  $R=1.54$ here. Given the simplicity of the models, the agreement is fairly good and we may conclude that the field effects reported here within both treatments are plausible and consistent.
		
		\end{document}